\documentstyle[floats,twocolumn,pra,aps]{revtex}
\input psfig.sty

\newcommand{\be}{\begin{equation}}
\newcommand{\ee}{\end{equation}}
\newcommand{\bea}{\begin{eqnarray}}
\newcommand{\eea}{\end{eqnarray}}
\newcommand{\bsa}{\begin{subeqnarray}}
\newcommand{\esa}{\end{subeqnarray}}

\newcommand{\beqa}{\begin{eqnarray}}
\newcommand{\beq}{\begin{equation}}
\newcommand{\eeqa}{\end{eqnarray}}
\newcommand{\eeq}{\end{equation}}

\newcommand{\dbyd}[2]{{\partial{#1}\over\partial{#2}}}

\newcommand{\dbydd}[2]{{\partial^2{#1}\over\partial{#2}}}

\newcommand{\pnt}{\>.}

\begin{document}

\title{A nonlinear evolution equation for sand ripples based on 
geometry and conservation}

\author{Zoltan CSAH\'OK$^{1,*}$, Chaouqi MISBAH$^1$, and Alexandre VALANCE$^2$}
\address{$^1$Laboratoire de Spectrom\'etrie Physique, Universit\'e Joseph Fourier 
(CNRS),
Grenoble I, B.P. 87, Saint-Martin d'H\`eres, 38402 Cedex, France}
\address{$^2$ Groupe Mati\`ere Condens\'ee et Mat\'eriaux, UMR 6626,
Universit\'e de Rennes 1, 35 042 cedex Rennes, France}
\address{$^*$ Permanent  address:  MTA Res. Inst. for Technical Physics and Materials Science, POBox 49, H-1525 Budapest, Hungary}

\author{(\today)}
\author{\parbox{397pt}{\vglue 0.3cm \small
From geometry and conservation we derive two 
nonlinear evolution equations for sand ripples.
In the case of a strong wind leading to a net erosion of the sand bed, ripples
obey the Benney equation. This leads either to order or disorder depending on whether
dispersion is strong or weak.
In the most frequent case where erosion is counterbalanced by deposition, we derive
a new one-parameter  nonlinear equation. It reveals
ripple structures which
then undergo a coarsening process at long times, a process which
then slows down dramatically with the growth of the ripple wavelength.
\\
\\
PACS numbers: 81.05 Rm
\\
}}
\maketitle

Ripples on sand in desert and see are fascinating 
patterns\cite{Bagnold41}. Despite
the fact that sand, and granular media in general, are very
familiar, in principle, to anyone, the understanding of their 
static and dynamical properties
still continues to pose a formidable challenge
to theoretical modeling\cite{Bideau}. A continuum description, as  those well
established for simple fluids (Navier-Stokes), or solids (Hookes law),
is lacking, most likely due to a strong interaction of disparate
scales. Whatever complex a continuum theory might be, it should
be compatible with symmetries and conservation laws. The present paper
deals with derivations of two generic nonlinear equations
for sand ripples based on geometry and conservation.

A model describing 
the physical origin of ripple formation has been presented
in the early forties by Bagnold\cite{Bagnold41}. This
model is based on energetic saltating grains impacting  the 
ripple. Since then, several contribution both 
analytical\cite{Anderson87,Hoyle97}
and 
numerical\cite{Anderson90,Nishimori93a,Nishimori93b,Landry94,Anderson93}
have allowed further elucidation 
of the problem. A derivation of a continuum
generic
nonlinear evolution equation of sand front is lacking, however.
This means in particular 
that the question of whether ordered, or disordered pattern,
would prevail is to date unanswered.  It is the main goal
of this
Letter to address these  questions   on the basis 
of geometry and conservation.
In the most interesting case where erosion is counterbalanced
by deposition on the average, we show here that
the generic equation in one dimension
(when the ripple is translationally invariant in the $y-$ direction)
takes the following form close to the instability threshold
\beq
{\partial h\over\partial t} =-{\partial^2 h\over\partial x^2} 
-\nu {\partial^3 h\over\partial x^3}-{\partial^4 h\over\partial x^4 }+
{\partial^2\over \partial x^2}[({\partial h\over\partial x})^2]
\label{MV}
\eeq
where $h$ is the ripple profile (in a dimensionless form)
which is a function of $x$ and $t$. As seen below this equation can
always be reduced to  one dimensionless parameter denoted here as $\nu$.
This equation reveals ripples. The wavelength
is first dominated by that of the linearly most
unstable mode. At long time they  coarsen before revealing
a dramatic slowing down of the wavelength increase. 
Note that the equation is local, which is a consequence of the
proximity of the instability threshold (see below).

For ease of presentation, we consider a one dimensional front. 
The most natural way of representing a front 
is to use intrinsic coordinates, 
namely the curvature $\kappa (s ) $ as a function
of the arclength $s$. Each point of the front is labelled by its vector 
position ${\bf r} (\alpha\; , t)$, where $\alpha$ is a time-independent
parametrization of the curve which can be taken at liberty
to lie in the interval  $0$ to $1$.  We obtain for
the evolution of the arclength $s$\cite{Misbah98}
\be
\dbyd{s}{t} = v_t\lbrack s(\alpha)\rbrack - v_t\lbrack s(0) \rbrack +
\int_{0}^{s} ds\prime \kappa v_n \pnt
\label{arc2}
\ee
This is simply obtained by setting $ds=\sqrt {g} d\alpha$, where $g$ is the induced
metric, and integrating $s=\int \sqrt {g} d\alpha$ by parts. Here we have
used the relation $dg/dt=2g ( \dbyd{v_t}{s} + \kappa v_n )$, where
$v_t$ and $v_n$ are the tangential and normal
velocities respectively.
In order
to derive the evolution equation for $\kappa$, we first need to 
evaluate the commutator  $\lbrack {d \over dt}, {\partial \over \partial s} \rbrack$.
We obtain
\beq
\lbrack {d \over dt}, {\partial \over \partial s} \rbrack=-(\dbyd{v_t}{s} + \kappa v_n)\dbyd{}{s} \pnt
\label{commu}
\eeq
Applying this identity to   $\theta$ (the angle between
the vertical axis and the normal vector),      we arrive at
\beq
\dbyd{\kappa}{t}= -( \dbydd{}{s^2} +\kappa^2) v_n + v_t \dbyd{\kappa}{s} \pnt
\label{courbure}
\eeq
Equations~(\ref{arc2}) and (\ref{courbure}) constitute
the evolution equations for the arclength and the curvature.
These equations are  general and only geometrical
concepts are evoked. 

The tangential velocity is a gauge. Indeed if one considers the front at some time
and its state at later time 
there is no way which allows us to state if a point $a$ 
(lying on the curve)
at time $t$ has becomes point $a'$ at $t+\Delta t$, or $a''$ (obtained 
by displacing $a'$ tangentially). 
The tangential velocity 
is fixed once the parametrization of the curve
is given\cite{Misbah98}). 
In contrast, $v_n$ is a physical quantity. 

A
way of viewing that the tangential velocity disappears
from the evolution equation is to work at given $s$, and not at  given $\alpha$.
Using (\ref{arc2}), the curvature evolution equation (\ref{courbure})
takes the form
\beq
{\partial \kappa \over \partial t}|_s=-[{\partial^2\over \partial s^2} +\kappa
^2 ]v_n -{\partial \kappa\over \partial s}\int _0^s ds' \kappa v_n
\label{kappa}
\end{equation}
Gauges usually introduce nonlocality.
Similar  formulations
were used in other contexts\cite{Foerster87,Kessler88}.
This equation constitutes  a general
nonlinear equation for curvature dynamics if
the normal velocity is known. This formulation
is powerful in numerical studies. If $v_n$
is known as a function of geometry 
then Eq. (\ref{kappa}) is a 
closed equation for curvature dynamics. Below we shall
write down the expression for $v_n$ inferred 
from symmetries and conservation. We shall then
use Eq. (\ref{kappa}) as a  starting point of our derivation
of evolution equations for the  ripples in the weakly
nonlinear regime  in terms of  
Cartesian coordinates where the front is represented
 by 
$z=h( x, t)$. 

Let us first illustrate our analysis on a situation that leads
to a well known equation in the literature. This will
serve later to determine  relevant
nonlinearities. For a rotationally invariant
system, the normal velocity must necessarily be  a function of only
those quantities which  are invariant under any surface
reparametrization. In one dimension the only quantity
which is intrinsic is the curvature and its odd derivatives
with respect to the arclength.
Generically,
the normal velocity has thus the form
\begin{equation}
v_n=C+a_1\kappa + a_2\kappa^2 + a_3 \kappa^3 +b_1 \kappa_{ss}+....
 \label{v4}
  \end{equation}
  From now on most of 
  differentiations will be subscripted for brevity.  Before
  proceeding further an important remark is in order
  concerning the assumption of locality in Eq. (\ref{v4}). For sand
  ripples discussed below, there
  are two kinds of grains
that contribute to the development of ripples
\cite{Anderson87,Hoyle97,Anderson90}: 
  the saltating ones (traveling on length
  scale $l_s\sim \lambda$, where $\lambda$ is the ripple wavelength)
  that have high kinetic energy, and the the low-energy splashed 
  grains traveling in reptation (or hopping)
  on a scale $a$ which is several (typically $6-10$) times 
smaller than $\lambda$.
  The saltating grains
are accelerated by the wind and this provides
the driving force that governs the motion of the surface in aeolian
ripple growth and in ripple translation.
In the most frequent case where erosion is counterbalanced
  by deposition, the population  of saltating grains remains
  almost constant (as recognized
  already by Bagnold\cite{Bagnold41} and in 
  \cite{Anderson87,Hoyle97,Anderson90}).
The saltating grains
serve merely to bring energy into the system,
the saltating population exchanges almost no grains with
the reptating population.
The information on the surface profile is propagated only
by reptating grains. Saltating grains loose their memory in the course of
their flight due to collisions and the turbulent airflow.
It follows then that $a$ (reptation
  length)
is the natural candidate\cite{Anderson87,Hoyle97,Anderson90} for
a characteristic length scale of interaction.
  Since $a/\lambda \ll 1$,
  the local assumption is legitimate. The ratio $a/\lambda$ 
  serves here as the small expansion parameter.

The constant $C$ in Eq. (\ref{v4}) expresses the fact that the front advances on the average at constant
velocity. For a straight front the velocity is $C$ (precisely as what
happens when a front is collecting particles from outside when exposed to
a given flux). For a weakly curved front we have
approximately $\kappa \simeq -h_{xx}$. Using this
in (\ref{v4}) and upon substitution  into (\ref{kappa})
we obtain 
\begin{equation}
h_t=C-a_1h_{xx}-b_1 h_{xxxx} +{C\over 2} h_x^2 \; ,
\label{ks}
\end{equation}
where we have truncated the expansion to leading order, as explained below.
First  we can have
$a_1>0$ or $a_1<0$. In the former case
the straight front is unstable. This is what happens
in a large number of situations (see Ref. \cite{Misbah94a}). 
The above derivation  has concentrated on the situation close enough to the
threshold where $a_1$ is small ( $a_1$ changes
sign upon variation of some
control parameter in a given system). 
For the truncation to make sense, 
$a_1$ must be
of order $\epsilon$ (some appropriate small quantity). This
 is  the case  sufficiently close
 to the instability threshold. For
the fourth 
 derivative
 to be of the order of the second one, we
  must have $\epsilon h_{xx}\sim  h_{xxxx}$, this implies
   in Fourier space  that $q$ (the wavenumber)
   must be of order $\sqrt {\epsilon}$. In other words
our equation is expected to be valid for slowly varying modulations on the scale
     of the typical length of interest in a given problem. This means
that modulations in physical units occurs on scales of order $1/\sqrt {\epsilon}$. Now in order
       that the linear part counterbalance the nonlinear one, we must have $\epsilon h_{xx}\sim h_{x}^2$. Since
        $x\sim 1/ \sqrt {\epsilon}$, this implies that $h\sim  {\epsilon}$. Using the same
  argument one arrives at the fact that time scales as $1/\epsilon^2$ (or equivalently
   the frequencies of interest are of order $\epsilon^2$).
    Once the scalings are known, it is a simple matter to show
     that
      other permissible nonlinearities 
      (e.g., $h_{xx}^2$) are of higher order contribution 
      (i.e., of order $\epsilon^4$).     

Note that the constant term in Eq. (\ref{ks}) 
is unimportant since it can be absorbed
on the l.h.s. upon a transformation $h\rightarrow h-Ct$. Note also (see below) that $b_1$ is generically
positive in order to ensure a well behaved solution at a short scale.
The sign of $C$ is however unimportant since changing it would simply correspond to
the transformation
$h\rightarrow -h$. The equation can be made free of parameter upon
appropriate rescaling.
Equation (\ref{ks})
is known under the name of Kuramoto-Sivashinsky(KS)\cite{Kuramoto76,Sivashinsky77}.
It models pattern formation
in different contexts (see Ref. \cite{Misbah94a}).
For large enough extent of the front in the $x$ direction,  it 
reveals spatio-temporal chaos.
If $a_1<0$ ,  there is no instability, and there would
thus be no need to keep the fourth derivative in Eq. (\ref{ks}).
Adding a small stochastic
force to $C$ (like shot noise in
Molecular Beam Epitaxy --MBE), 
we obtain the well known Kardar-Parisi-Zhang\cite{Kardar86} equation, introduced
to model kinetic roughening in MBE. 

Having  illustrated our study on a reference example, we are now in 
a position to deal with 
ripple formation under wind blow. On the one hand
the front is not advancing on the average, in principle. Thus
the constant $C$ must be set to zero  (this is unimportant as seen below).
On the other hand, the wind causes
the normal front velocity to be orientation-dependent.
We first concentrate  on the situation where there is a strong
erosion, so that the front is surrounded by an atmosphere of flying
grains (a sort of reservoir). The front motion can thus
globally loose or gain grains from the atmosphere, so that
no constraint must be imposed. In the presence of 
anisotropy (due to the wind) the most
natural way is to write
\beqa
v_n=&&a_1 \kappa +a_2 \kappa^2 + b_1 \kappa_{ss}
+\alpha_1 \sin \theta + \alpha_2 (\sin \theta)^2 \nonumber \\
&&  + \beta_1
 \frac{\partial^2 }{\partial s^2}( \sin \theta )
+ \beta_2
 \frac{\partial^2 }{\partial s^2}( \sin^2  \theta )
  + \gamma_1 \kappa \sin \theta +......\pnt \label{v7}
  \eeqa
  The terms in $\sin \theta$ express the fact that the growth velocity depends
  on the local slope of the front. In the present case, the direction
  perpendicular to the $z$ axis (from which $\theta$ is measured) is favored.
  Had we wanted to 
  give a greater importance to the $x$ direction, we would then
  have expanded $v_n$ in 
  power of $\cos \theta$. However, it is a simple matter
  to realise that this is unimportant for our purposes.
  In the limit of a weakly curved front, the substitution
  of (\ref{v7}) into (\ref{kappa}) yields to leading
  order
  \beq
  h_t=-\alpha_1 h_x- a_1h_{xx} -\beta_1 h_{xxx}-b_1h_{xxxx}+\alpha_2 h_x^2
  \label{beney}
  \eeq
  This equation is known in the literature under the name of 
  Benney equation\cite{Benney66}.
  It has been derived recently from a microscopic model in the context
  of step-bunching dynamics during
  sublimation of a vicinal surface\cite{Sato95,Misbah96}. The same
  equation arises in other contexts such
  as phase dynamics for traveling modes\cite{Janiaud92},
  and in some models of traffic flow\cite{Hong95}. The first derivative
  term can be absorbed in the temporal derivative by means of a Galilean transformation (i.e. $x\rightarrow
  x-\alpha_1 t$). The scaling of space, time and amplitude with
  $\epsilon$ are obviously the same
 as for the KS equation.
 The third derivative (not present in the KS equation)
  is of higher order 
  $1/\epsilon^{1/2}$ if all the scales in the KS part are set 
  to one (as seen
  before space scales as $1/\sqrt{\epsilon}$, and $h\sim \epsilon$,
  so that $a_1 h_{xx}\sim h_{xxxx}\sim h_x^2 \sim \epsilon^3$, while
  $h_{xxx}\sim \epsilon^{5/2}$). 
Thus the expansion would make sense in principle only if $\beta_1$ were small enough (of order $\sqrt{\epsilon}$),
a demand whose realization depends on the system under consideration.
This apparent
difficulty can be circumvented by noting that
  $h_{xxx}$ contributes
  to the imaginary part of the linear  growth rate (if we seek  a solution in the form $e^{iqx+\omega t}$, where
  $\omega$ is the growth rate), which concerns thus propagative terms, whereas $h_{xx} $ and $h_{xxxx}$
  produce real contributions. 
  Thus one can 'split' time into a slow part corresponding
  to growth of perturbations, and a fast part corresponding to propagation.

  Upon rescaling of the space, time and amplitude,
  the Benney equation can be rewritten in a form in which only one parameter
  survives. Therefore all the coefficients can be set to unity except one,
  let say $\beta_1$. Depending on the strength of
  this coefficient, the
  dynamics is either chaotic  for
  $\beta_1 $ of order or smaller than  one (we recover the KS dynamics), or exhibits a rather ordered
structure drifting sideways for $\beta_1$ of order few unities\cite{Misbah96}. 

For sand ripples,
it seems that in  most  situations an equilibrium between
erosion and deposition sets in due to the drag force of the transported grains
exerted on the wind
(the greater the number of transported grains is, the weaker
the wind gets and the less it erodes the sand bed). In other words, the
number of transported grains remains constant in average; there is
neither net erosion  nor deposition. In order to treat
that case we should impose the global conservation condition.
In order to ensure this  all
the homogeneous terms in (\ref{v7}) must be left out, except
the linear terms in $\kappa$ and $\sin \theta$. Indeed  
the area $A$ bounded by the front and some horizontal axis behaves
in course of time as
$\partial A/\partial t=\int ds v_n$, and as global
conservation imposes $\partial A/\partial t=0$, all terms giving a 
non-zero contribution
to that area should be removed (the
sand front moves because either particles have left 
the region of interest, or other grains have landed from
a neighboring part). In this case,
the front dynamics is governed to leading order by 
\beq
h_t=-\alpha_1 h_x- a_1h_{xx} -\beta_1 h_{xxx}-b_1h_{xxxx}+\beta_2
\frac{\partial^2}{\partial x^2}(h_x^2)
\label{MV3}
\eeq
For an instability $a_1$ must be positive, $b_1$ must be positive as well in order
to introduce a short wavelength cut-off, while the sign of $\beta_2$ is 
unimportant, since it can be changed upon the transformation $h\rightarrow -h$.
Obviously $\alpha_1 h_x$ can be absorbed in $h_t$ via a Galilean transformation,
and the sign of $\beta_1 $ is unimportant as well.   Space and time scale
with $\epsilon$ as in the KS and Benney equations, while the scale of $h$
here is of order one, in a marked contrast
with the KS limit. This scaling
has an important consequence, to be discussed below.
What makes the other higher contribution small 
is precisely the scaling of space (higher and higher derivatives are of
smaller and smaller contributions). After rescaling (and absorbing
$h_x$ in $h_t$) only one parameter survives
and the {\it principal  equation for ripples} can be written in a canonical form
(Eq. (\ref{MV})).
The linear dispersion relation of Eq. (\ref{MV})
takes the form (by looking for perturbations
in the form $h\sim e^{iqx +\omega t}$) $\omega=q^2-q^4 - i\nu q^3$. 
The structureless
state is linearly unstable against perturbations with wavenumbers smaller  than
$q_c=1$. A physical  origin  of the
instability was put forward long time ago by
Bagnold\cite{Bagnold41}. Here we assume
that the threshold has been reached, and thus
we take a negative sign in front of 
the second derivative. 
Numerical solutions for  sizes
$L\ge 10 \lambda_c=2\pi$
of  Eq.~(\ref{MV}) reveals an evolution towards a steady state 
with a given wavelength. In a marked  contrast to the KS equation which exhibits
spatiotemporal chaos (or Benney equation when dispersion
is small), the new evolution equation (\ref{MV}) leads to  ripple
pattern 
(Fig.1). The wavelength is first close to that of the most
dangerous mode. At long time, the structure coarsens producing thereby
wider and wider dunes\cite{remark}. Then the coarsening
slow down dramatically. This  feature
agrees  with experiments\cite{Anderson90}. 
An extensive study will be presented in the future. 
\begin{figure}
\centerline{
\psfig{figure=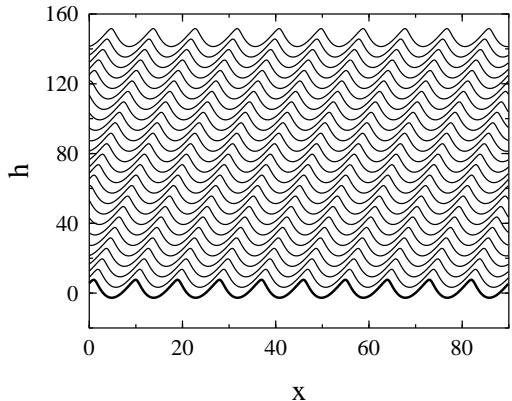,width=8cm}
}
\caption{The ripple profile at different time. The consecutive snapshots
have been shifted upward to show the drift.}
\end{figure}

Finally it is important to make some important
comments.  (i) Since $h$ is of order one,
while $x$ is of order $1/\sqrt{\epsilon}$, the slope 
is of order  $\sqrt{\epsilon}$, and thus remains small (see Fig.1).
Had the  slope been of order one
(as in some nonlinear equations\cite{Misbah98}), avalanches
would manifest themselves, and no steady-solutions would have been
possible. (ii) Figure 1 corresponds to equation (\ref{MV}).
As stated above changing the sign  in front of the nonlinear term corresponds
to making an up-down operation on Figure 1.  Inspecting
several examples of sand  ripples conveys the strong impression
that it is the situation in Figure 1 which seems likely. 
(iii) Since $h\sim 1$ and $\lambda\sim 1/\sqrt{\epsilon}$,
the ratio of the amplitude to wavelength close to the instability
threshold scales as ${\epsilon}^{1/2}$. Thus close to threshold
the amplitude is several times smaller than  the wavelength
(usually ripples have an amplitude which is  approximately
10 times smaller
than 
their
wavelength).
(iv) We have
deliberately, for sake of simplicity, considered a one dimensional 
structure (that is the ripples are translationally invariant
along the $y$ direction). The extension to two dimensions is feasible,
and is currently under investigation\cite{Csahok98}. This should be crucial
with regard to the study of possible secondary instabilities of ripples.

In  summary we have derived a generic  nonlinear equation
to describe sand ripple dynamics. The study is based on geometry and
conservation.
While apparently close equations (e.g., the  KS equation) lead 
to spatiotemporal
chaos, the new equation reveals rather steady ripples which coarsen with
time. 
The advantage of this study lies in the fact that no matter how
complex the physics might be, the equations close to the instability
threshold
must be of the sort given here as long as symmetries and conservations
are preserved.
In turn, geometry and conservation can not, by their very nature, provide
the values of coefficients 
The same holds for the existence of instability.
Thus there is a need in future to derive our equation 
starting from a given  'microscopic' model and to determine the 
dimensionless coefficient $\nu$ in terms of  physical quantities.
Other remarks are in order. We have limited ourselves to leading order, which
is valid close enough to the instability point\cite{remark}.
Expansion to higher order is 
straightforward. 
We have assumed that the normal velocity is both local in 
space and in time.
This is valid close to the instability threshold. If reptation
length $a$ remains the relevant length scale, we may expect
our assumption of locality to hold at arbitrary
distance from threshold, since $a/\lambda$ is the small
parameter of the expansion.
Finally, it goes 
without saying that
the present study can have impact on other systems than the sand
ripples.


\begin{thebibliography}{10}
\bibitem{remark}
The next term is given by $a_1[h_x^3]_x$. Its smallness is due to that of $a_1$
  which is small close to the threshold. One of its effects is to delay the
  coarsening process.
\end{thebibliography}
\end{document}